\documentclass[twocolumn,twoside]{article}

\newcommand{\beq}{\begin{equation}}
\newcommand{\eeq}{\end{equation}}
\newcommand{\ctft}[1]{\ensuremath{\mathcal{#1}}}    %Continuous Fourier Transform
\newcommand{\cfv}{\ensuremath{\Omega}}              %Continuous Fourier Transform variable
\newcommand{\dtft}[1]{\ensuremath{\hat{#1}}}        %Discrete Time Fourier Transform
\newcommand{\dfv}{\ensuremath{\omega}}          %Discrete Time Fourier Transform variable
\usepackage{epsfig}
\usepackage{graphicx}
\usepackage{graphics}

\begin{document}
\newtheorem{definition}{\it Definition}
%\markboth{IEEE Transactions on Signal Processing, Vol.xxx, No.xxx, xxxx}
%{Vasiloglou and Maragos: FIF Spectrum}

%\setcounter{page}{0}

\title{Spectrum of Fractal Interpolation Functions \thanks{Submitted: Sep. 2002.
This  research work  was supported by the Greek Secretariat for
Research and Technology and by the European Union under the
program E$\Pi$ET-98 with Grant \# 98$\Gamma$T26, when both authors
where with with the Department of Electrical \& Computer
Engineering, National Technical University of Athens. Nikolaos
Vasiloglou is now with Analytics 1305 LLC Atlanta Georgia USA. Petros Maragos is  with the Department of Electrical \&
Computer Engineering, National Technical University of Athens,
Zografou, 15773 Athens, Greece. Email:
\texttt{nvasil@ieee.org,maragos@cs.ntua.gr}}}

\author{
{\Large
Nikolaos Vasiloglou, \emph{Member, IEEE}
and
Petros Maragos, \emph{Fellow, IEEE}
 }
  }

\maketitle

\begin{abstract}
In this paper we compute the Fourier spectrum of the Fractal Interpolation Functions FIFs as introduced by Michael Barnsley. We show that there
is an analytical way to compute them. In this paper we attempt to solve the inverse problem of FIF by using the spectrum
\end{abstract}

%\section{Introduction}

%%%%%%%%%%%%%%%%%%%%%%%%%%%%%%%Iterated Function Systems%%%%%%%%%%%%%%%%%%%%%%%%%
\section{Iterated Function Systems}
\begin{figure}[!htb]
\centerline{ \psfig{figure=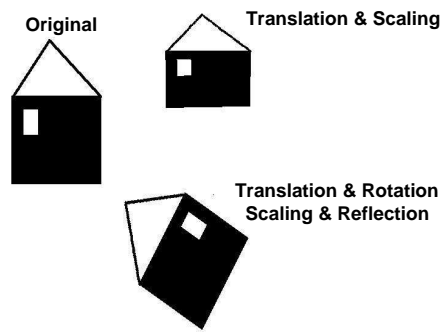,height=5cm}}
\caption{An affine transform can  translate rotate and flip a
2-dimensional shape. } \label{houseaffine}
\end{figure}

The affine transform performs translation stretching and rotation
on a given set. In the special case of two dimensions the affine
transform on a set $\mathcal S$ in 2-D space is described by the
equation:
\[
w(x,y)=\left[%
\begin{array}{cc}
  a & b \\
  c & d \\
\end{array}%
\right]
\left[%
\begin{array}{c}
  x \\
  y \\
\end{array}%
\right] +
\left[%
\begin{array}{c}
  e \\
  f \\
\end{array}%
\right]
\]
where $(x,y)\in {\mathcal S}$. The effects of an affine transform
on a set are depicted in fig.~\ref{houseaffine}. The union of N
affine transformations is called the Hutchinson operator:
$W=\bigcup_{n=1}^{N}w_n$. For a specified metric the distance
$h({\mathcal A,B})$ between  two sets ${\mathcal A,B}$ can be
defined. Under certain conditions \cite{barnsley} the Hutchinson
operator is contractive, \linebreak $h(W({\mathcal A}),W({\mathcal
B}))\leq s h({\mathcal A,B}),\quad s<1$. Successive iterations
with Hutchinson operator on a random set results in a sequence of
a sets that converges in the \textit{attractor} of the operator
$\mathcal A$, which satisfies the condition $ {\mathcal
A}=W({\mathcal A})=\bigcup_{n=1}^{N}w_n({\mathcal A})$. Any system
that uses the Hutchinson operator in order to generate iteratively
the attractor $\mathcal A$ is called \textit{Iterated Function
System} (IFS).

%%%%%%%%%%%%%%%%%%%%%%%%%%%%%%%Fractal Interpolation Functions%%%%%%%%%%%%%%%%%%%%%%%%%%%%%%%%%%%%%%%%%
\subsection{Fractal Interpolation Functions}
Fractal Interpolation Functions (FIF) is a special case of  the
2-dimensional IFS and maintain all their characteristics. FIF
attractors are continuous functions that can be used to model
continuous signals. FIF interpolate a given set of $N+1$ points
$(x_{n},y_{n}),\quad n=0,1\dots N$
\[x_0<x_1<x_2<x_3<\dots<x_N\]
The FIF that interpolates the above set is comprised of  $N$ affine maps:
\[
w_{n} \left[ \begin{array}{c} x \\ y \end{array} \right] =\left[ \begin{array}{cc}
a_{n} & 0 \\
c_{n} & d_{n}
\end{array} \right] \left[ \begin{array}{c} x \\ y \end{array} \right]
+ \left[ \begin{array}{c} e_{n} \\ f_{n} \end{array} \right] ,
n=1,\dots,N .\]

The necessary condition is that the Interpolation Function passes
from the $N+1$ initial points,

\begin{equation}
w_{n}\left[ \begin{array}{c} x_{0} \\ y_{0} \end{array} \right]
 =\left[ \begin{array}{c} x_{n-1} \\ y_{n-1} \end{array} \right]
\mbox{ and }
w_{n}\left[ \begin{array}{c} x_{N} \\ y_{N} \end{array} \right]
 =\left[ \begin{array}{c} x_{n} \\ y_{n} \end{array} \right] \quad ,
\label{conditions}
\end{equation}
\[
n=1,\dots,N
\]

We call order of the FIF, the number $N$ of the affine maps. The
conditions provide 4 equations for 5 parameters , so  $d_n$, the
vertical scaling factor is chosen to be the free parameter. If we
solve the above equations for $a_n,c_n,e_n,f_n$ in terms of $d_n$,
we find :
\begin{eqnarray}
a_{n} &=& \frac{x_{n}-x_{n-1}}{x_{N}-x_{0}},\label{a} \\
e_{n} &=& \frac{x_{N}x_{n-1}-x_{0}x_{n}}{x_{N}-x_{0}},\\
c_{n} &=& \frac{y_{n}-y_{n-1}}{x_{N}-x_{0}} - \frac{d_{n}(y_{N}-y_{0})}{x_{N}-x_{0}},\label{c} \\
f_{n} &=& \frac{x_{N}y_{n-1}-x_{0}y_{n}}{x_{N}-x_{0}}-d_{n}\frac{x_{N}y_{0}-x_{0}y_{N}}{x_{N}-x_{0}}. \label{f}
\end{eqnarray}

Let the real numbers $a_n,c_n,e_n,f_n$ be defined by
(\ref{a}-\ref{f}). Barnsley \cite{barnsley} introduced the
operator $T$ for the class $\mathcal{C}$ of continuous functions,
$T:\mathcal{C}\rightarrow \mathcal{C}$ by
\[
(TF)(x)=c_{n}\ell_{n}^{-1}(x)+d_{n}F(\ell_{n}^{-1}(x))+f_{n}\quad
\]
\[
\forall x\in [x_{n-1},x_{n}]\;\;,n=1,2\dots,N ,
\]
 and
\[
\ell_{n}:[x_{0},x_{N}]\rightarrow [x_{n-1},x_{n}]
\]
 the invertible transformation
\[
\ell_{n}(x)=a_{n}x+e_{n}.
\]
\label{T}

The above operator:
\begin{itemize}
\item  is a contraction mapping according to \textit{Hausdorff}
metrics, \item satisfies the conditions of (\ref{conditions}),
\item has a unique fixed point, the function $F\in\mathcal{C},$
 \[ (TF)(x)=F(x) ,\quad \forall x\in[x_{0},x_{N}]. \]
\end{itemize}
The following restrictions guarantee the contractivity of $T$
operator:
\begin{itemize}
\item $\; a_{n}\;<1$ \item $|d_{n}|<1.$
\end{itemize}
Barnsley's operator is very useful because the attractor of an FIF
can be generated with the iterative application of $T$ on an
initial signal  $F(0)$ : $F_{m+1}=TF_{m}$,
\[F=\lim_{m\rightarrow\infty}T^{\circ m}(F_0), F_0 \in \mathcal{C} \]

\begin{figure}[htb]
\centerline{ \psfig{figure=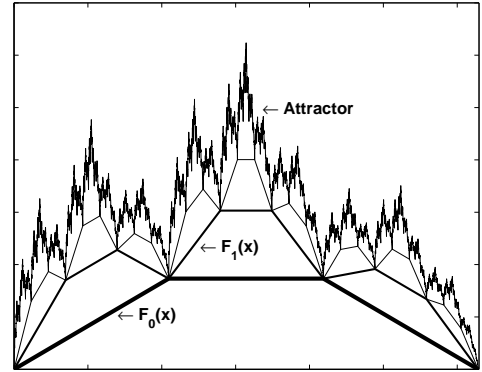,height=5cm}}
\caption{Formation of the FIF attractor after successive
iterations} \label{fifsteps}
\end{figure}

%%%%%%%%%%%%%%%%%%%%%%%%%%%%%%%%DISCRETE CASE%%%%%%%%%%%%%%%%%%%%%%%%%%%%%%%%%%%%%%%%%%%%%%
\subsection{Discrete FIF}
\begin{figure}[!htb]
\centerline{ \psfig{figure=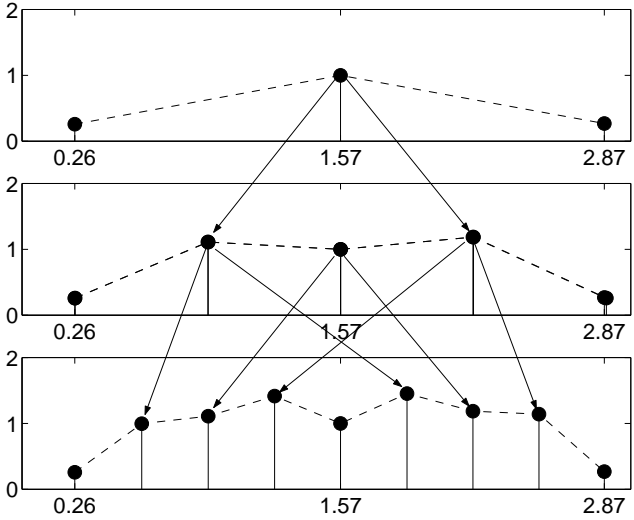,height=7cm}}
\caption{Generation of a discrete FIF.} \label{dfifsteps}
\end{figure}

One of the most interesting properties of the FIF is that its
attractor is independent of the initial signal (initiator). If the
initiator is a continuous signal, for example the linear
interpolation between the given interpolation points, all the
instances throughout all the steps of the iterations will be
continuous signals, fig.~\ref{fifsteps}. On the other hand if the
initiator is a single point the attractor formed after infinite
iterations will be a continuous signal, but  signals instances of
the FIF throughout the iterations will be discrete signals,
fig.~\ref{dfifsteps}. Although during all the iterations the
instances are discrete signals or strictly mathematically speaking
finite countable sets, the attractor is a continuous signal, an
infinite and uncountable set. Instances of a the formation of an
FIF from a discrete initiator are shown in fig.~\ref{dfifsteps}.

It is essential to show that a good choice of the initiator is the
$N+1$ given interpolation points $(x_{n},y_{n}),n=0,1\dots N$.
After the first iteration in each of the $N$ subintervals between
the points, $N-2$ new points are generated fig.~\ref{dfifsteps}.
Let these points be $(x_s,y_s), s=1,\dots,(N-1)N$. Notice that all
these $N+1+(N-2)N=N^2-N+1$ points belong to the attractor of FIF.
This wouldn't be true if the initiator was a different set apart
from the given interpolation points. By repeating this procedure
after $m$ iterations we get an $M(m)$-point discrete sequence that
is a sampling of the FIF's attractor, with sampling period
$T_s=\frac{1}{M}$. If the initiator included any other irrelevant
point, that would not be mapped to an attractor's point after a
finite number of iterations. It can be proved that if the
initiator is not the set of the initial interpolation points, then
the error of the formed sequence after the $m$th iteration from
the attractor decreases and goes to zero as $m\rightarrow\infty$.
The number of the attractor's samples is :
\begin{eqnarray}
N+1&\; \nonumber \\ \nonumber
N+1+(N-1)N &=& N^2+1 \\ \nonumber
N+1+(N-1)N+(N-1)N^2 &=& N^3+1 \\ \nonumber
\vdots & & \vdots \\
M=M(m)=N^m+1.
\label{SamplingRate}
\end{eqnarray}
The discrete signal formed after the $m$th iteration over the
discrete initiator, is called discrete FIF :

\begin{equation}
f[n]=F(x_n), \quad n=0,1,\dots,M
\end{equation}

Before expanding Barnsley operator ${\mathcal T}$ introduced above
for the discrete FIF it is necessary to make clear  that according
to the strict definition of Fractals the discrete FIFs are not
fractal sets because they are finite. Discrete FIFs must be
considered as  approximations of the continuous.

Since the Barnsley operator assumes infinite resolution, it
doesn't apply in discrete signals. So for discrete FIF the
following modified operator is used:

\begin{eqnarray}
{\mathcal T} f[n] =\sum_{k=1}^{N} (d_k {\uparrow}^N f[\frac{n-e_k}{a}] & + c_k(\frac{n-e_k}{a})+f_n)  \\ \nonumber
(u(n-e_{k-1})-u(n-e_k)). &
\label{Tdiscrete}
\end{eqnarray}

\begin{figure}[!htb]
\centerline{ \psfig{figure=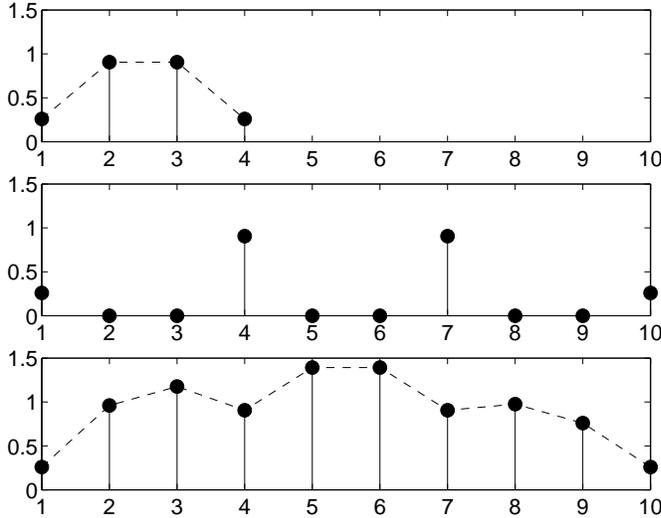,height=7cm}}
\caption{Barnsley operator for discrete signals. The initiator is
upsampled and interpolated according to (\ref{Tdiscrete} }
\label{dbarnsleyoper}
\end{figure}

The Symbol ${\uparrow}^N f[n]$ defines that within two successive
samples of $f[n]$, $N-1$ zeros have been interpolated,
fig.~\ref{dbarnsleyoper}. The term $f_n$ denotes the parameter of
FIF as defined in (\ref{f}) and should not be confused with the
discrete signal $f[n]$.

Similarly to the  continuous case, a discrete  FIF of m iterations
can be constructed with the following procedure:

\begin{equation}
f_{m+1}[n]={\mathcal T} f_m[n].
\end{equation}

\begin{figure}[!htb]
\begin{center}
\setlength{\unitlength}{.9in} \thicklines
\begin{picture}(5.3,1.8)(.4,0)
\put(.8,1.4){\vector(1,0){.55}} \put(1.4,1.4){\vector(1,0){.6}}
\put(2,1.2){\framebox(.4,.4){\shortstack{$\uparrow N$}}}
\put(2.4,1.4){\line(1,0){.4}}
        \put(2.8,1.4){\vector(1,0){.6}}
\put(2.8,1.4){\line(0,-1){.7}} \put(2.8,0.7){\line(-1,0){.4}}
\put(2,0.5){\framebox(.4,.4){\shortstack{${\dtft q}(\dfv)$}}}
\put(2,0.7){\line(-1,0){.6}} \put(1.4,.7){\vector(0,1){.65}}
\put(1.4,1.4){\makebox(0,0)[c]{\huge $\bigoplus$}}
\put(.7,1.5){\makebox(0,0)[l]{${\dtft g}(\dfv)$}}
\put(3.2,1.5){\makebox(0,0)[l]{${\dtft f}(\dfv)$}}
\end{picture}
\vspace*{-.5in}
\end{center}
  \caption{Block diagram of the FIF Genaration.
   \label{Blockdiag}  }
\end{figure}
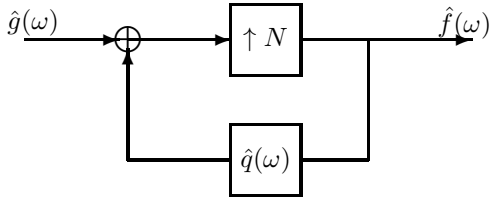

It is obvious that when $m\rightarrow\infty$ the discrete FIF
becomes continuous. Notice that ${\mathcal T}$ depends on the
number of iterations $m$. More specifically the parameters of FIF
as defined in (\ref{a}-\ref{f}) depend on the $x_n$. The $x_n$
change value because of the upsampling. The generation of an FIF
can be represented in terms of a linear system as shown in
fig.~\ref{Blockdiag}.

%%%%%%%%%%%%%%%%%%%%%%%%%%%COMPUTATION OF FIFS's FOURIER SPECTRUM%%%%%%%%%%%%%%%%%%%%%%%%%%%%%%
\section{Computation of FIF's Fourier Spectrum}
In order to simplify (\ref{a}-\ref{f}), it is very
convenient to adopt the following assumptions.
\begin{itemize}
    \item $x_{0}=0,\quad x_N=1,$
    \item $ F_{0}=F_{N}=0.$
    \item and the points are evenly spaced.
\end{itemize}

Then(\ref{a}-\ref{f}) becomes,

\begin{eqnarray}
a_{n} & = & \frac{1}{N},  \\
e_{n} & = & \frac{n-1}{N},  \\
c_{n} & = & F_{n}-F_{n-1},  \\
f_{n} & = & F_{n-1}.
\label{Affine Parameters}
\end{eqnarray}
The above equations show that the FIF parameters are decoupled
from each other. This is very important because they can be
estimated independently. Moreover it is clear that if $a_n$
parameter is estimated then all parameters can be found directly,
except for $d_n$.

\begin{figure}[htb]
\centerline{ \psfig{figure=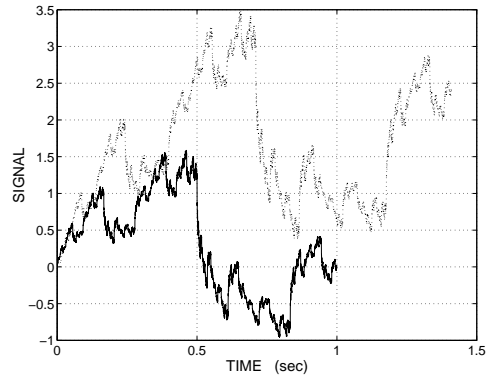,height=5cm} } \caption{ An
example of FIF out of range $[0,1]$. Applying an affine transform
we can tie it at points $(0,0)$ and $(1,0)$. } \label{fif1}
\end{figure}

Any FIF can be transformed to an equivalent FIF that satisfies the
first two conditions without loosing its fundamental properties.
More specifically fig.~\ref{fif1} shows how a given FIF can be
transformed so as to satisfy the conditions for the first and the
last point. It is convenient to use an auxiliary affine map
$w_{aux}$ that will rotate, scale and translate the given one. The
$w_{aux}$ transform is invertible and does not affect the
intrinsic parameters $a_n,d_n$.

Let
  $A=\{(x,y):F(x)=y\},$ be the original attractor and

\begin{equation}
{\bigcup}_{n=1}^{N} w_n (A)=A
\label{TransFIF}
\end{equation}
where $w_n$,
\[
w_{n} \left[ \begin{array}{c} x \\ y \end{array} \right] =\left[ \begin{array}{cc}
a_{n} & 0 \\
c_{n} & d_{n}
\end{array} \right] \left[ \begin{array}{c} x \\ y \end{array} \right]
+ \left[ \begin{array}{c} e_{n} \\ f_{n} \end{array} \right] , n=1,\dots,N
\]

The new transformed attractor is $A'=\{(x',y'),F'(x)=y' \}.$
The new points are connected with the initial
\[ \left[ \begin{array}{c} x' \\ y' \end{array} \right] =
\left[ \begin{array}{cc} a_{aux} & 0 \\ c_{aux} & d_{aux} \end{array} \right]
 \left[ \begin{array}{c} x \\ y \end{array} \right] +  \left[ \begin{array}{c} e_{aux} \\ f_{aux} \end{array} \right]
\]
That means $A'=w_{aux}A$. These new points belong to
a new FIF. Setting in (\ref{TransFIF})
$A=w^{-1}_{aux} A'$ and applying the map $w_{aux}$,

\begin{equation}
{\bigcup}_{n=1}^{N} w_{aux}w_n w^{-1}_{aux}(A')=A',
\label{fif_equivelance}
\end{equation}
with \[
w^{-1}_{aux}=\left[
\begin{array}{cc}
1/a_{aux} & 0 \\
-\frac{c_{aux}}{a_{aux}d_{aux}} & 1/d_{aux}
\end{array}
\right].
\]
By setting
$f_{aux}=e_{aux}=0,$
$d_{aux}=1$, the affine maps of the new FIF are :

\begin{eqnarray}
w'_{n} \left[ \begin{array}{c} x' \\ y' \end{array} \right] =
\left[
\begin{array}{cc}
a_{n} & 0 \\
(c_n-d_n c_{aux}+c_{aux}a_n)/a_{aux} & d_{n}
\end{array}
\right]
&
\nonumber
\\ \nonumber
\left[ \begin{array}{c} x' \\ y' \end{array} \right]
+
\left[ \begin{array}{c} a_{aux}e_{n} \\c_{aux}e_n +f_n \end{array} \right] &
\\ \nonumber
\end{eqnarray}
Notice that the new FIF has the same order and the same $d_n$ parameters.

%%%%%%%%%%%%%%%%%%%%%%%%SPECTRUM OF CONTINUOUS FIF%%%%%%%%%%%%%%%%%%%%%%%%%%%%%%%
\subsection{Spectrum of Continuous FIF}
The application of the above simplifications to the Barnsley
operator $T$  results in the following equation for the FIF:

\begin{equation}
\begin{array}{l}
F(x)=\sum_{n=1}^{N}(d_{n}F(\frac{x-e_{n}}{a})+ \\
+c_{n}(\frac{x-e_{n}}{a})+f_n)(u(x-e_{n-1})-u(x-e_{n}))
\end{array}
\end{equation}
and
\[ u(x)=\left\{ \begin{array}{ll}
                   0 & \mbox{$x<0$} \\
           1 & \mbox{$x \geq 0$}
        \end{array}
    \right.
\]
$G(x)$  is the piecewise linear function between the interpolation
points :
\[ G(x)=\sum_{n=1}^{N}c_{n}(\frac{x-e_{n}}{a}+f_n)(u(x-e_{n-1})-u(x-e_{n})).\]

\begin{figure}[htb]
\centerline{ \psfig{figure=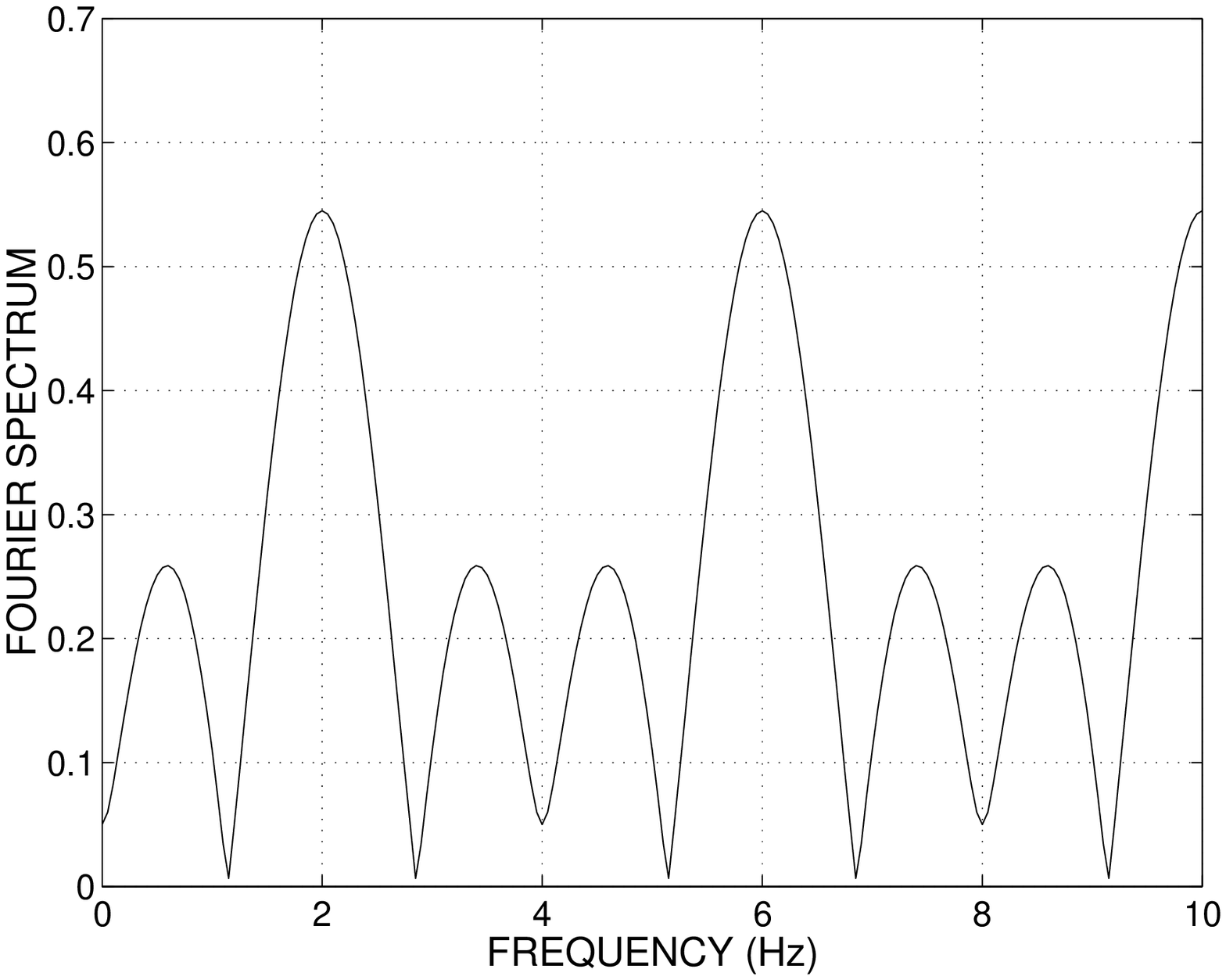,height=5cm} } \caption{ An
example of ${\ctft Q}(\cfv)$. } \label{Q(w)fig}
\end{figure}

Applying continuous fourier transform \cite{oppenheim}:

\[ {\ctft F}(\cfv)=\int_{- \infty}^{+ \infty} F(x) e^{-i \cfv x} dx \]

\begin{eqnarray}
F(x) & \longleftrightarrow & {\ctft F} (\cfv)  \nonumber \\
{\ctft F}(\cfv) & = & {\ctft G}(\cfv)+a{\ctft F}(a\cfv)\sum_{n=1}^{N}d_{n}e^{-i\cfv e_{n}},
\end{eqnarray}

\begin{equation}
\begin{array}{l}
{\ctft G}(\cfv)=\sum_{n=1}^{N}(\frac{e^{-i\cfv\frac{n-1}{N}}-e^{-i\cfv\frac{n}{N}}}{(i\cfv)^{2}}
(\frac{c_{n}}{a}+i\cfv(f_{n}-\frac{c_{n}e_{n}}{a})) \\
+\frac{c_{n}}{a}\frac{\frac{n-1}{N}e^{-i\cfv\frac{n-1}{N}}
-\frac{n}{N}e^{-i\cfv\frac{n}{N}}}{i\cfv})
\end{array}
\label{G(w)}
\end{equation}

We define the function :

\begin{equation}
{\ctft Q}(\cfv)=a\sum_{n=1}^{N} d_{n}e^{-i\cfv \frac{(n-1)}{N}}=a \sum_{n=0}^{N-1} d_{n+1} e^{-i \frac{\cfv n}{N}}
\label{Q(w)}
\end{equation}
Notice that the above function is the discrete time fourier
transform of the discrete sequence $\{d_1, d_2,\dots d_N \}$, so
we deduce that $Q(\cfv-2k\pi N)=Q(2(k+1)\pi N-\cfv),\quad
k=1,2,\dots\quad $ (fig.~\ref{Q(w)fig}).

\begin{figure}[htb]
\centerline{ \psfig{figure=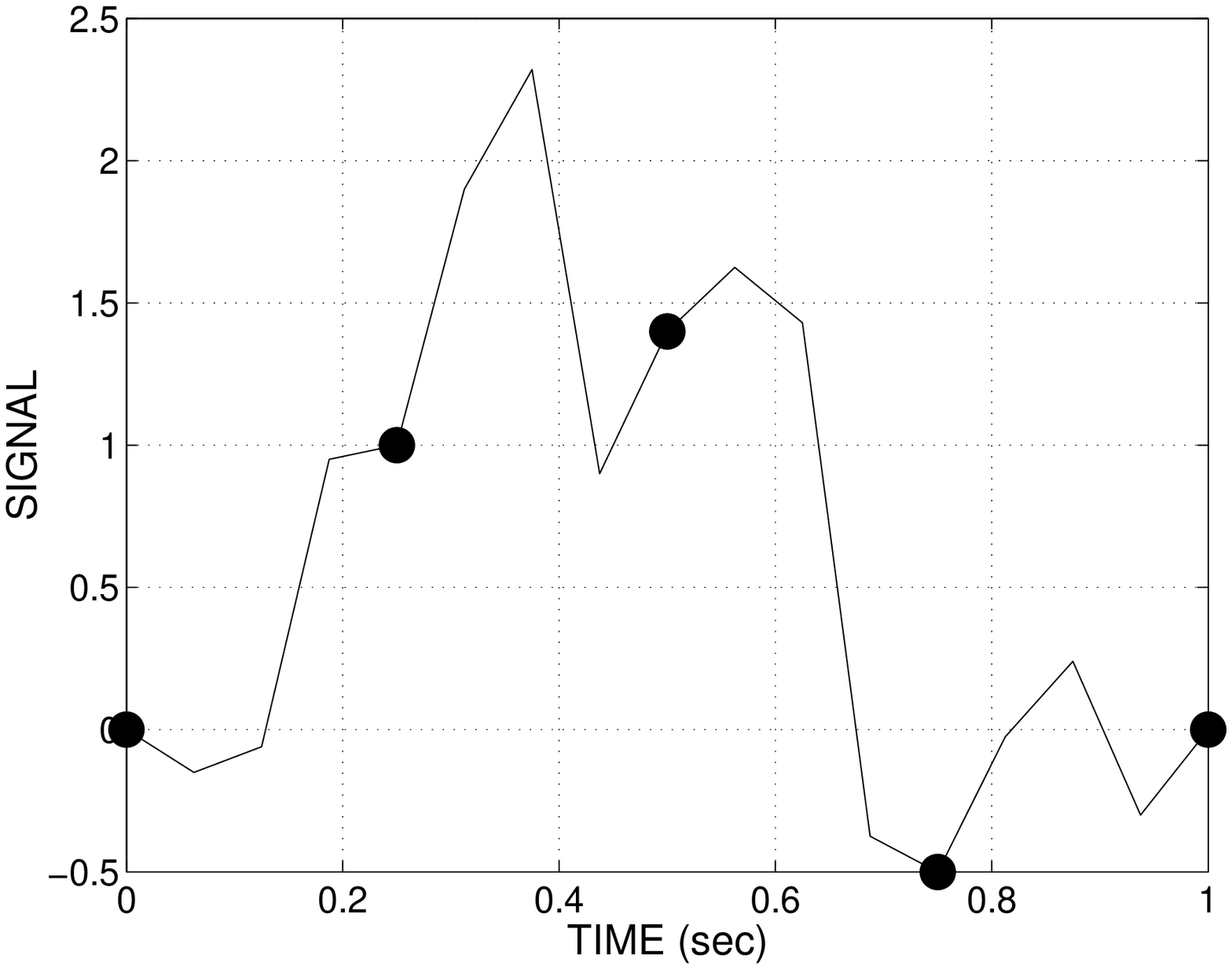,height=3cm}
\psfig{figure=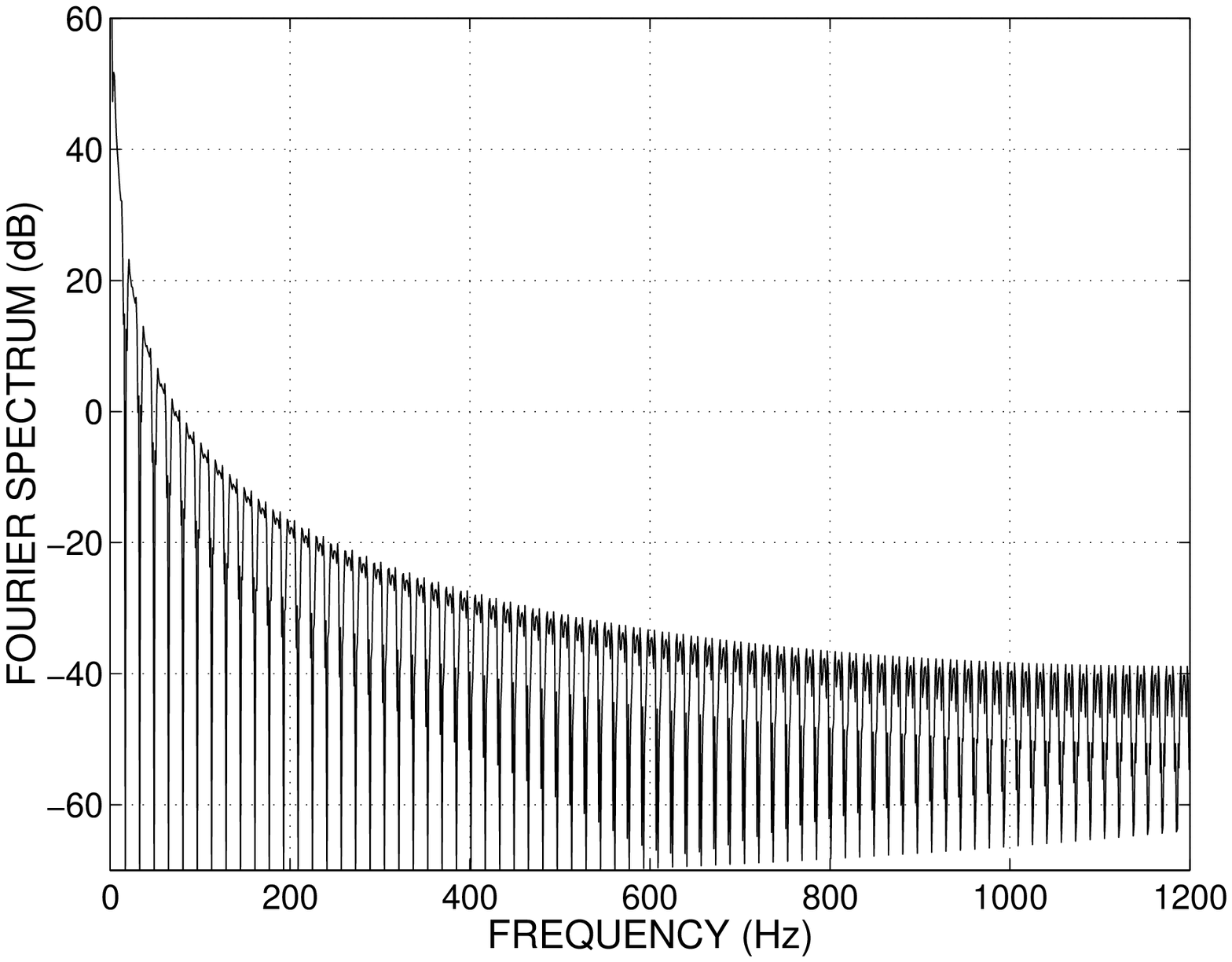,height=3cm} } \centerline{
\psfig{figure=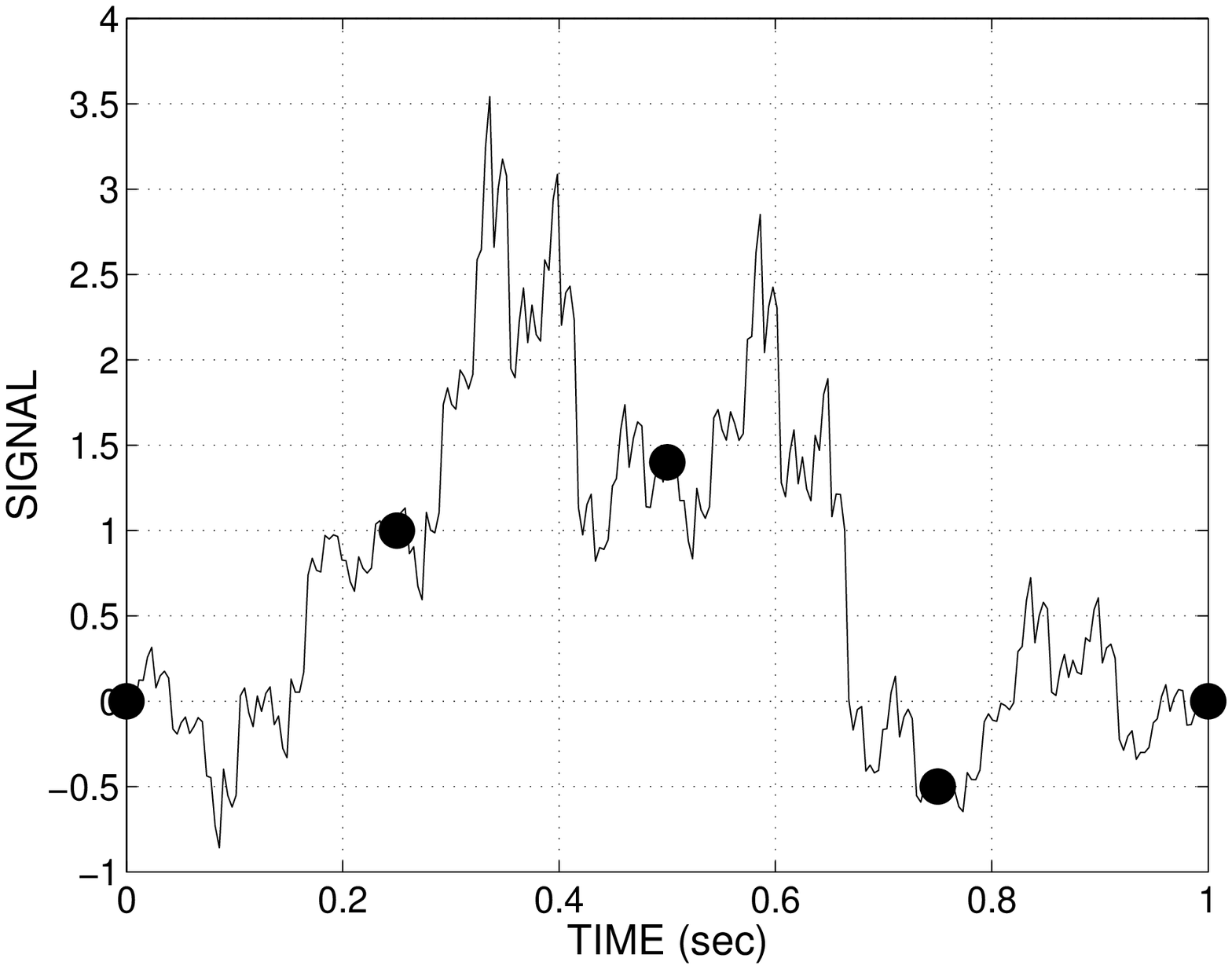,height=3cm}
\psfig{figure=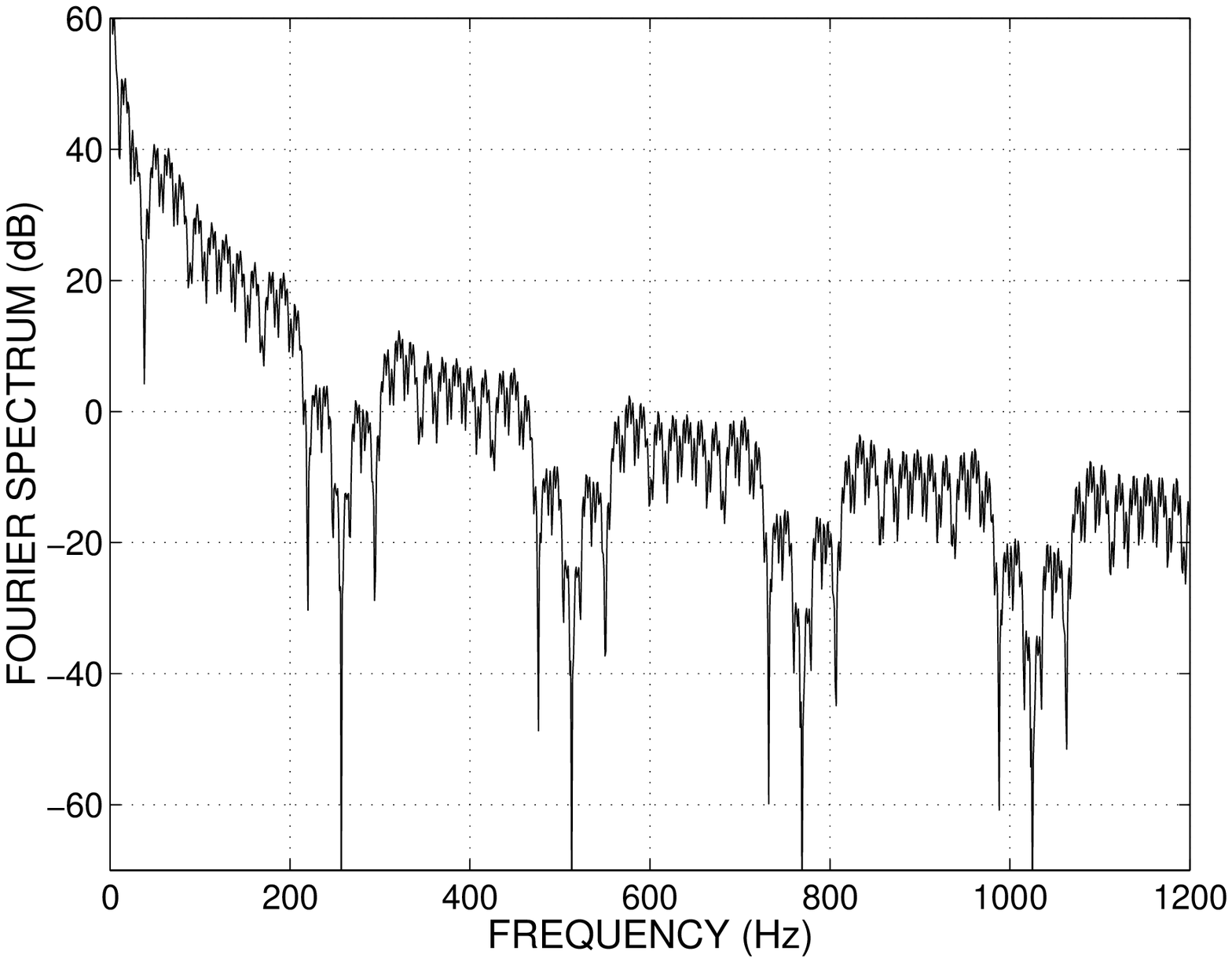,height=3cm} } \centerline{
\psfig{figure=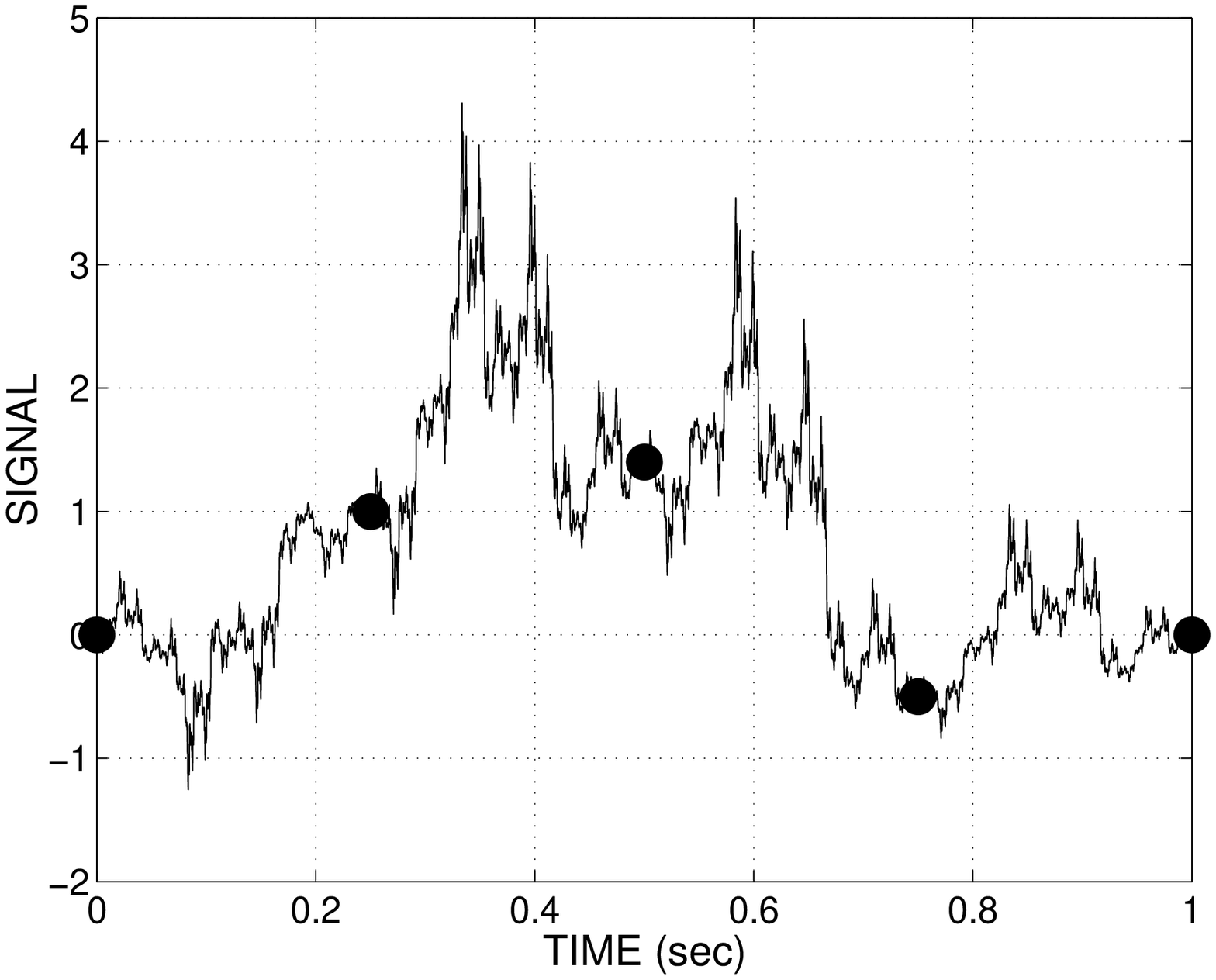,height=3cm}
\psfig{figure=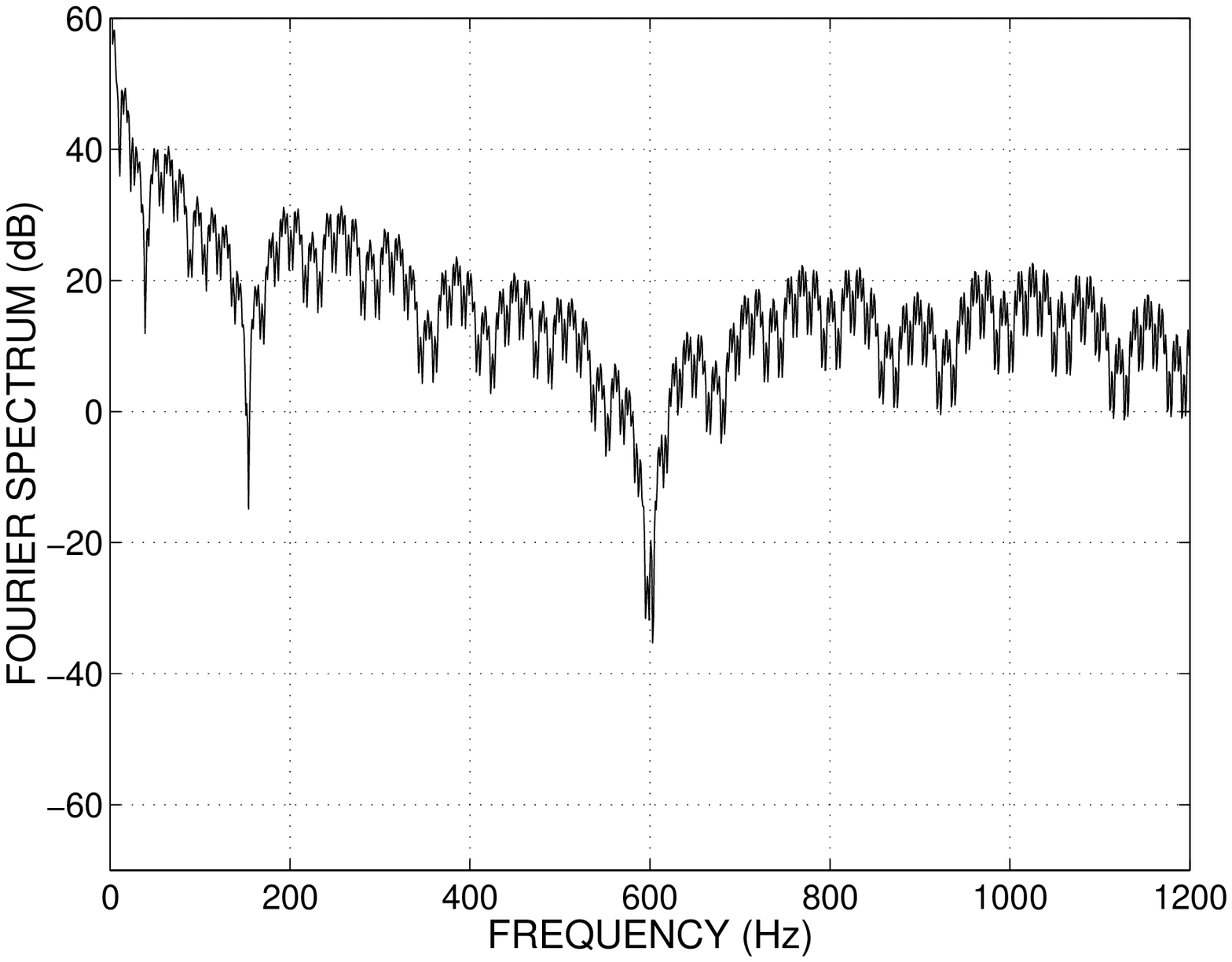,height=3cm} } \caption{ Left
column: Fractal interpolation between points $
(0,0),(0.25,1),(0.5,1.4),(0.75,-0.5),(1,0) $ after 1,3,5
iterations. right column: Corresponding spectrums. }
\label{specexample}
\end{figure}

The Fourier spectrum satisfies the following equation:
\begin{equation}
{\ctft F}(\cfv)={\ctft Q}(\cfv){\ctft F}(a\cfv)+{\ctft G}(\cfv)
\end{equation}

Through Barsley Operator in frequency domain the Fourier spectrum
can be computed iteratively, fig.~\ref{specexample}.
\begin{equation}
{\ctft F}_{m+1}(\cfv)={\ctft Q}(\cfv){\ctft F}_{m}(a\cfv)+{\ctft G}(\cfv)
\end{equation}
 ${\ctft F}_{0}(\cfv)=0.$
\\
After infinite iterations:
\begin{equation}
{\ctft F}(\cfv)=\sum_{i=0}^{\infty}{\ctft G}(\cfv a^{i})\prod_{j=0}^{i-1} {\ctft Q}(\cfv a^{j})
\label{FT}
\end{equation}

%%%%%%%%%%%%%%%%%%%%%%%SPECTRUM OF DISCRETE FIF%%%%%%%%%%%%%%%%%%%%%%%%%%%%%%%%%%
\subsection{Spectrum of Discrete FIF}
From the (\ref{FT}) it is obvious that the FIF signals are not
band-limited. As a result the spectrum of a discrete FIF is
aliased. Assume $T_s=\frac{1}{M}$ has been chosen as the sampling
period, then the Discrete Time Fourier Transform (DTFT)
\cite{oppenheim} is :
\[
{\dtft f}(\dfv)=\sum_{n=0}^{M} f[n]e^{-i\dfv n}, \dfv=\cfv T_s.
\]

As in continuous case for the computation of the DTFT the
${\mathcal T}$ operator is used in frequency domain:

\begin{equation}
\dtft{f}_{m+1}(\dfv)=\dtft{f}_m(a\dfv)\sum_{p=1}^{N}d_k a e^{-i\dfv e_p}+ \dtft{g}_m(\dfv)
\end{equation}

Posing the analogy between the continuous and the discrete case we
define the $q$ function :
\[
\dtft{q}_m(\dfv)=a\sum_{p=1}^{N}d_p e^{-i\dfv e_p}.
\]

\[  \dtft{q}(\dfv)=\sum_{p=1}^{N} d_p e^{-i{\dfv} e_p},\quad p=0,1,\dots,N.
\]
and $e_p=\frac{p-1}{N}(M-1),$ so
\begin{equation}
\begin{array}{l}
{\dtft q}(\dfv)=\sum_{p=0}^{N-1} d_{p+1} e^{\frac{-i{\dfv} p}{N}(M-1)} \\
=\sum_{p=0}^{N-1} d_{p+1} e^{(-i{\dfv} p)(N^{m-1}-\sum_{i=1}^{m-2}N^i)},\quad p=0,1,\dots,N.\quad
\end{array}
\end{equation}

\begin{equation}
{\dtft f}_{m+1}(\dfv)={\dtft q}_m(\dfv){\dtft f}_m(a\dfv)+ {\dtft g}_m(\dfv)
\end{equation}

After $m$ iterations
\begin{equation}
 {\dtft f}_m(\dfv)=\sum_{i=0}^{m}{\dtft g}_{m-i}(\dfv a^{i})\prod_{j=0}^{i-1} {\dtft q}_{i-j+1}(\dfv a^j)
\label{DFT}\\
\end{equation}

\begin{figure}[htb]
\centerline{ \psfig{figure=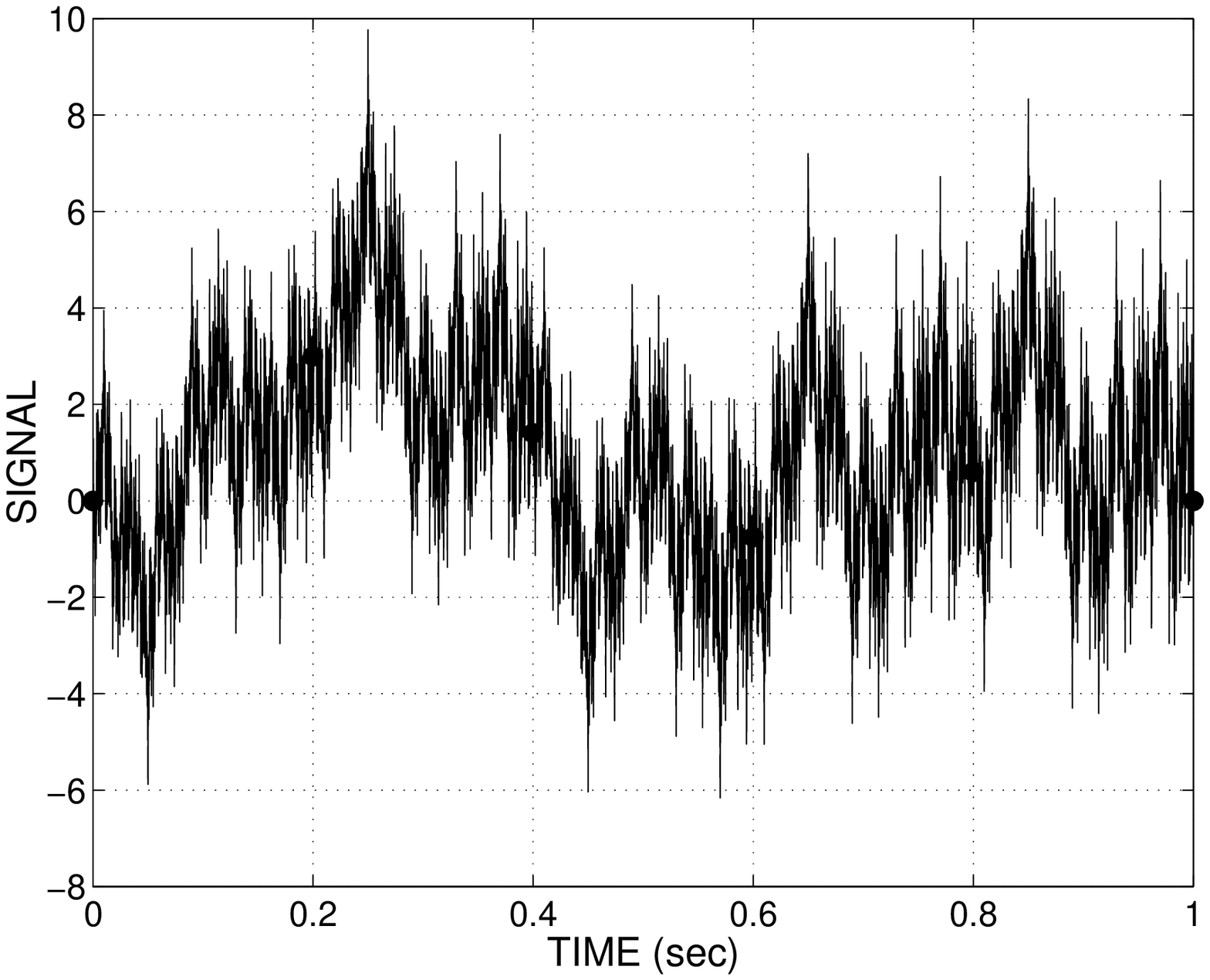,height=3cm}
\psfig{figure=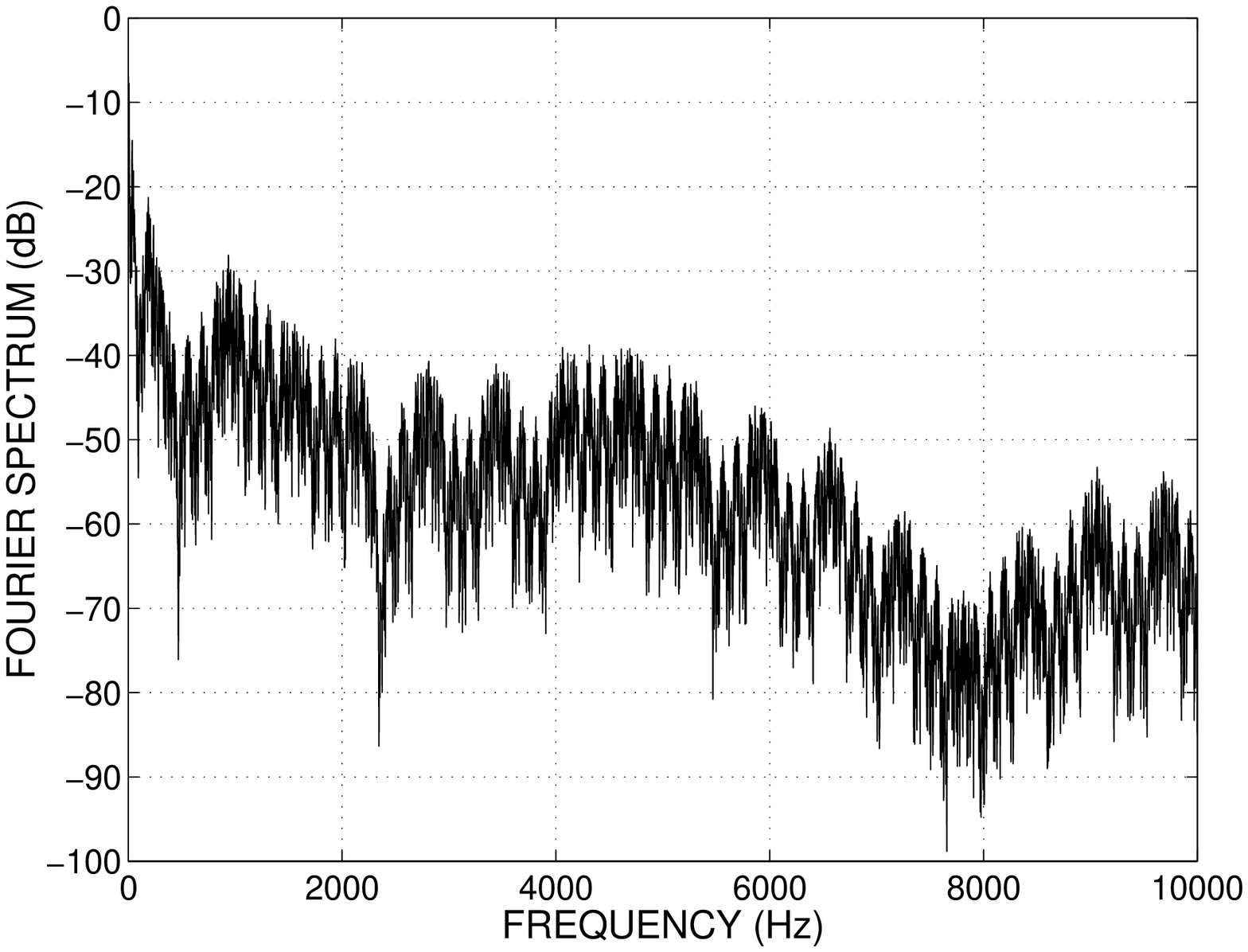,height=3cm} }
\caption{5-point FIF
attractor with $d_n=-0.74,0.8,-0.77,0.85,0.88. $ and the spectrum
after the 5th iteration. } \label{alias}
\end{figure}

The   Discrete Fourier Transform can be computed after keeping the
frequency values between $0$ and $2\pi$ and sampling the spectrum
at
$\dfv=0,\frac{2\pi}{M},2\frac{2\pi}{M},\dots,(M-1)\frac{2\pi}{M}$
, fig.~\ref{alias}.

In the continuous case ${\ctft Q}(\cfv)$ has period $2\pi N$. In
the discrete case where ${\dfv} \in [0,2\pi]$ the ${\dtft
q}(\dfv)$ function has period $2\pi\frac{N}{M-1}$.

%%%%%%%%%%%%%%%%%%%%%%%%%%%%%%%%%%%%%%%%%%%%%%%%%%%%%%%%%%%%%%%%%%%%%%%%%%%%%%%%%%%%%%%%%%%%%%%%%%%%%%%%%%%
%%%%%%%%%%%%%%%%%%%%%%%%%%%%%%%%%%%%%%%%%%%%%%%%%%%%%%%%%%%%%%%%%%%%%%%%%%%%%%%%%%%%%%%%%%%%%%%%%%%%%%%%%%%
\section{FIF parameter estimation using spectrum}
FIF modelling of signals has been proposed by Mazel \cite{mazel},
using information of signal in time domain. In this section the
spectrum of signal is used to estimate its parameters, provided
the FIF order is known.

Given the signal $f[n],n=0,1,\dots,M$ , let $N$ be  FIF's
estimated order. We expect its spectrum to satisfy the following
equation:

\[
{\dtft q}[\dfv_{k}]=\frac{{\dtft f}_{m+1}[\dfv_{k}]-{\dtft
g}[\dfv_{k}]}{{\dtft f}_{m}[a\dfv_{k}]},
\]
\[
 \quad
a\dfv_{k}=0,\frac{2\pi}{MN},2\frac{2\pi}{MN},\dots,(MN-1)\frac{2\pi}{MN}.
\]
All values that zero denominator are excluded. Notice that
$f_{m+1}[n]$ and $f_m[n]$ signals must have the same length.
Considering also that $f_{m+1}[0]=f_m[0]$ and
$f_{m+1}[M(m+1)]=f_m[M(m)]$, where $M(m),M(m+1)$ their lengths. In
order to equalize their lengths it is necessary to interpolate
with zeros $f_m[n]$. Because of the term ${\dtft
f}_{m}[a\dfv_{k}]$ the computation of FFT must be on
$\frac{2k\pi}{MN},k=0,\dots,NM-1$. This is done by padding the
signal with $MN-M$ zeros. The $d_n$ are determined by the solution
of the system of linear equations.
\begin{equation}
\left[
\begin{array}{c}
\dtft{q}(0)\\\dtft{q}[\dfv_{1}]\\\dtft{q}[\dfv_{2}]\\ \vdots\\\dtft{q}[\dfv_{N}]
\end{array}
\right]=
\left[
\begin{array}{ccccc}
1 &  1                       &\dots &1\\
1 & e^{\frac{-i\dfv_{1}}{N}} &\dots &e^{\frac{-(N-1)i\dfv_{1}}{N}}\\
1 & e^{\frac{-i\dfv_{2}}{N}} &\dots &e^{\frac{-(N-1)i\dfv_{2}}{N}}\\
\vdots&\vdots                &\dots       &\vdots\\
1 & e^{\frac{-i\dfv_{R}}{N}} &\dots &e^{\frac{-(N-1)i\dfv_{R}}{N}}\\
\end{array}
\right]
\left[
\begin{array}{c}
d_{1}\\d_{2}\\d_{3} \\ \vdots\\d_{N}
\end{array}
\right]
\end{equation}

%%%%%%%%%%%%%%%%%%%%%%%%%%%%%%%%%%%%%%%%%%%%%%%%%%%%%%%%%%%%%%%%%%%%%%%%%%%%%%%%%%%%%%%%%%%%%%%%%%%%%%%%%%%
%%%%%%%%%%%%%%%%%%%%%%%%%%%%%%%%%%%%%%%%%%%%%%%%%%%%%%%%%%%%%%%%%%%%%%%%%%%%%%%%%%%%%%%%%%%%%%%%%%%%%%%%%%%

\begin{figure}[htb]
\centerline{ \psfig{figure=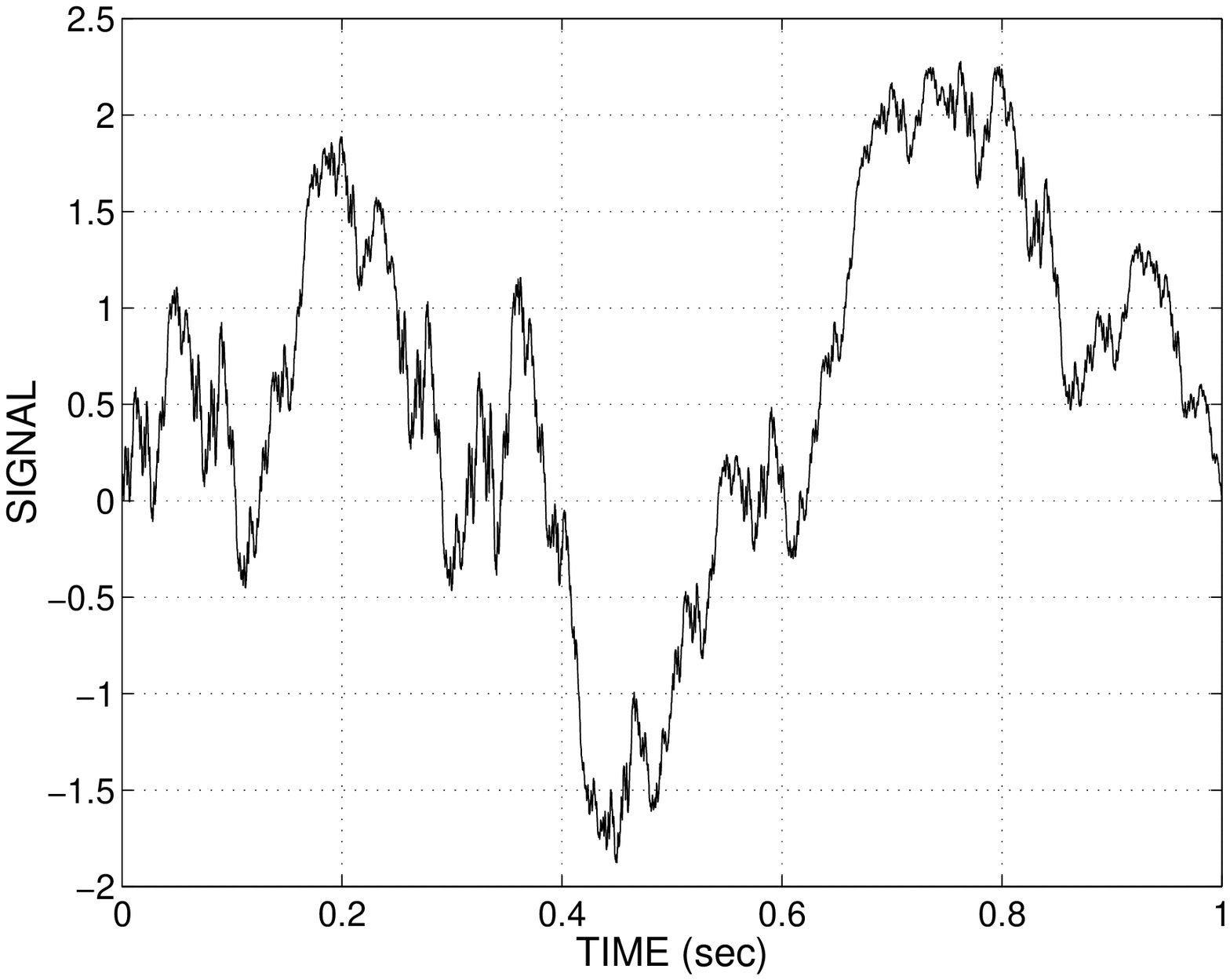,height=3cm} (a) }
\centerline{ \psfig{figure=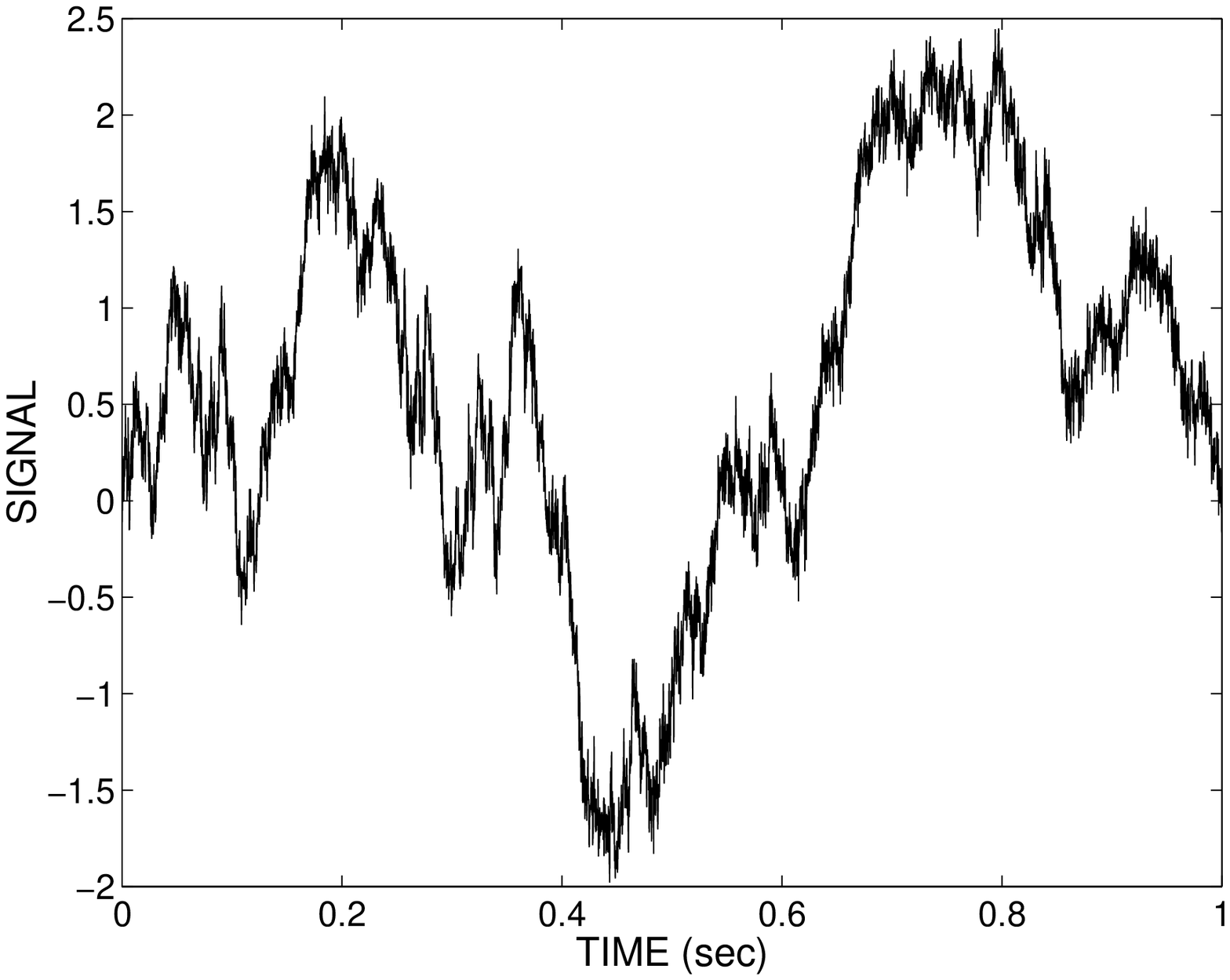,height=3cm}(b)
\psfig{figure=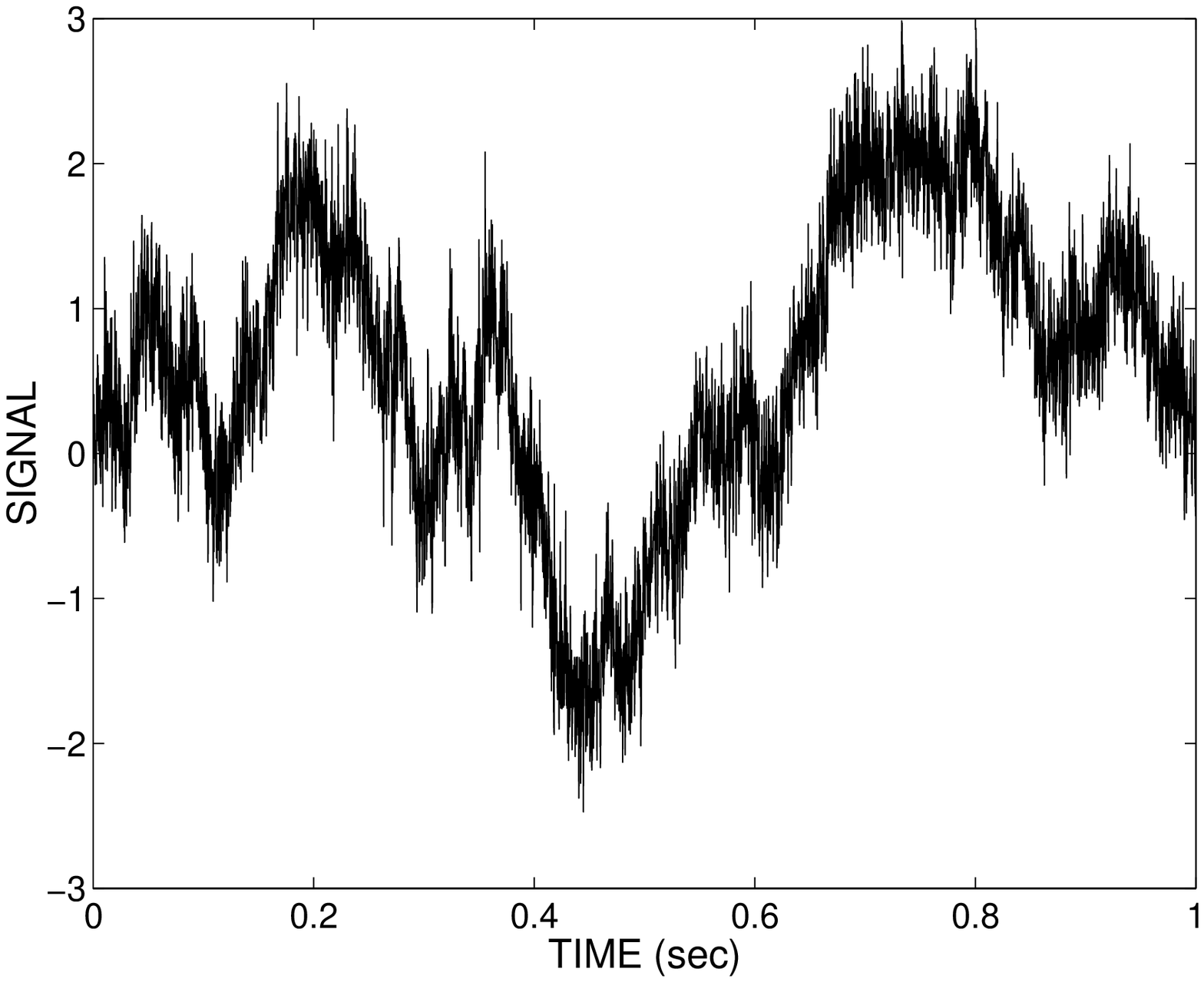,height=3cm} (c) } \centerline{
\psfig{figure=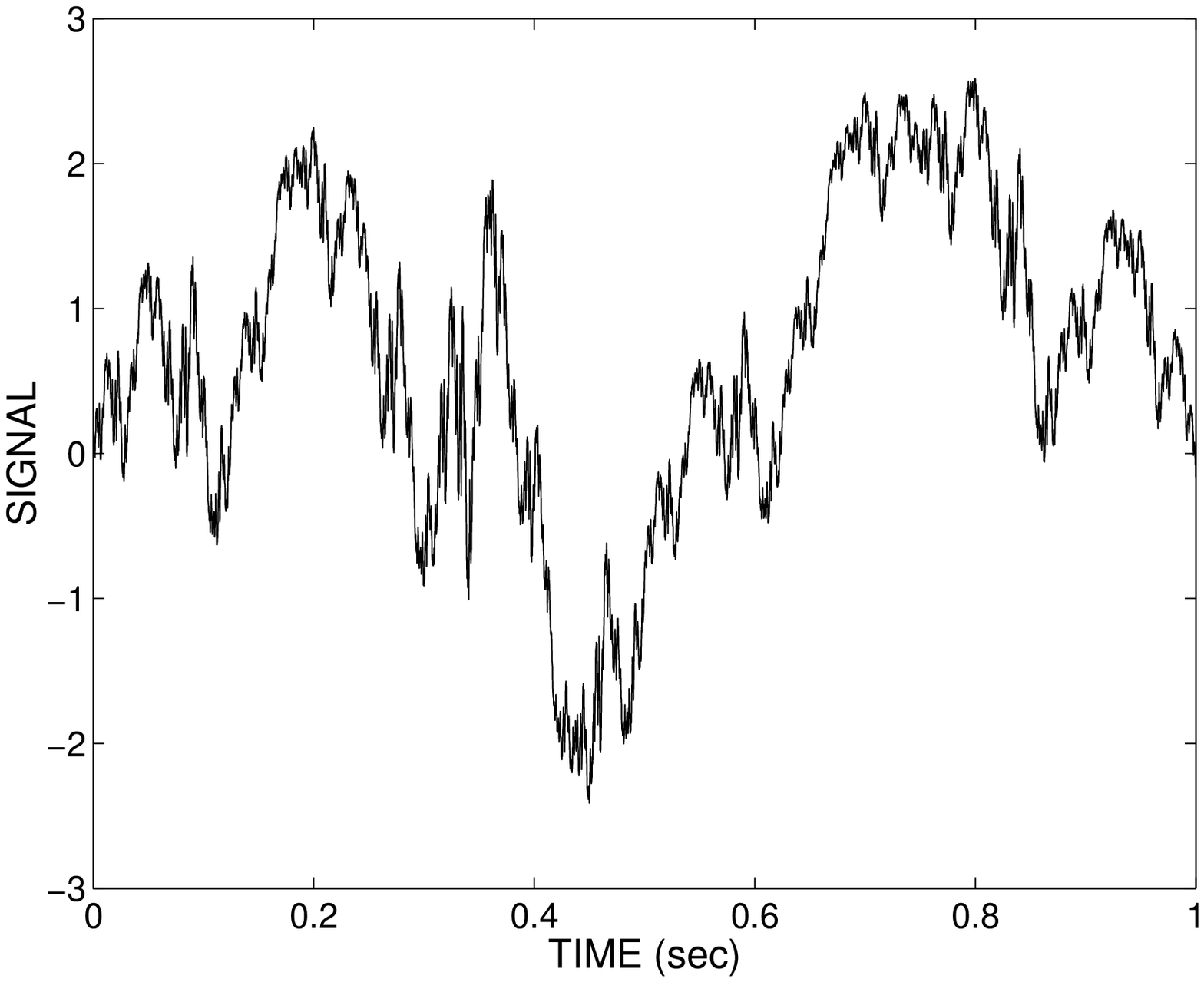,height=3cm} (d)
\psfig{figure=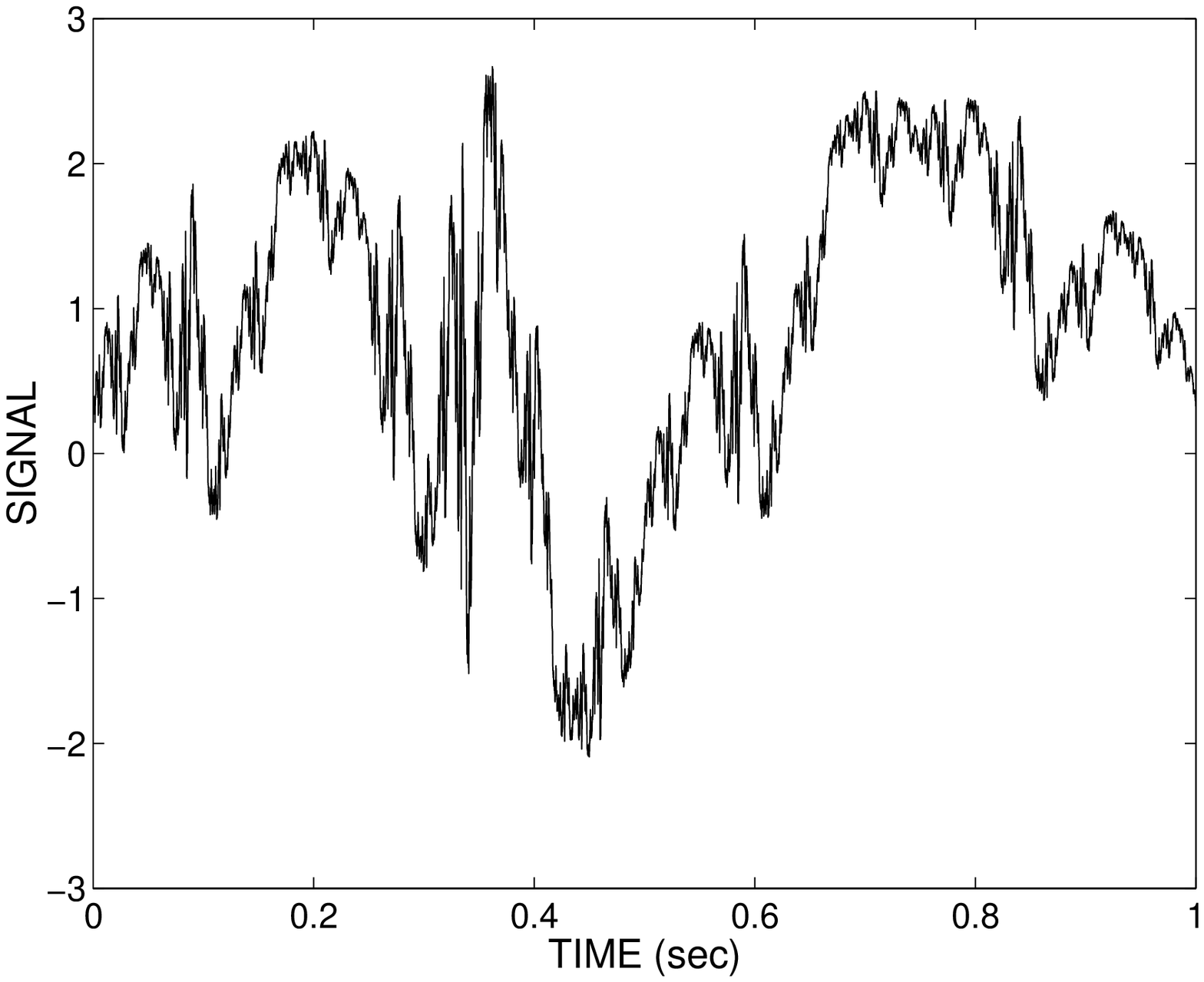,height=3cm} (e) } \caption{(a)
FIF without noise. In (b),(c) FIF contaminated with noise
$\mathit{SNR}=20$ and $\mathit{SNR}=10$. In (d),(e) reconstructed
FIF. } \label{noisy_fif}
\end{figure}

\section{Results}
\begin{figure}[htb]
\centerline{
\psfig{figure=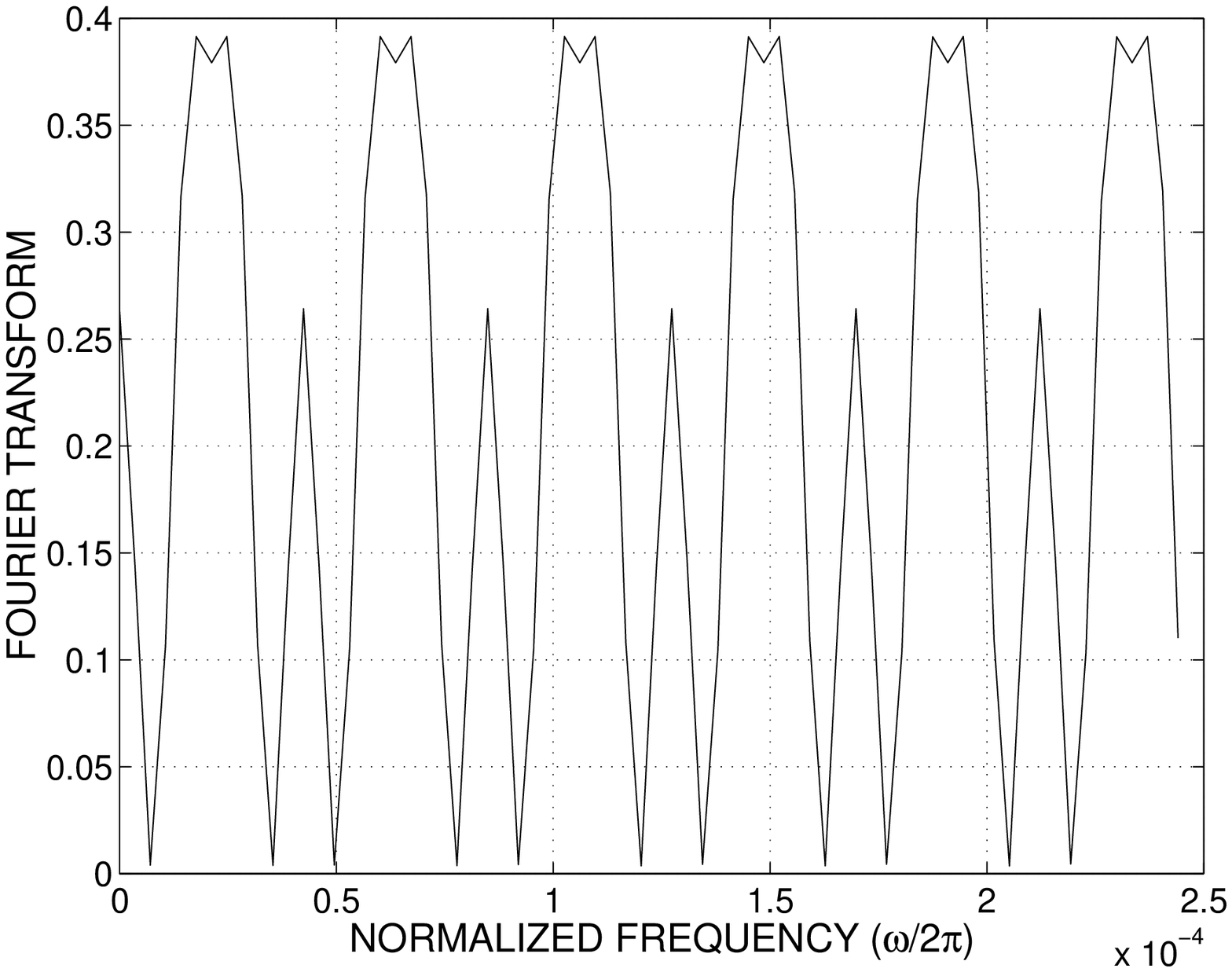,height=3cm}
\psfig{figure=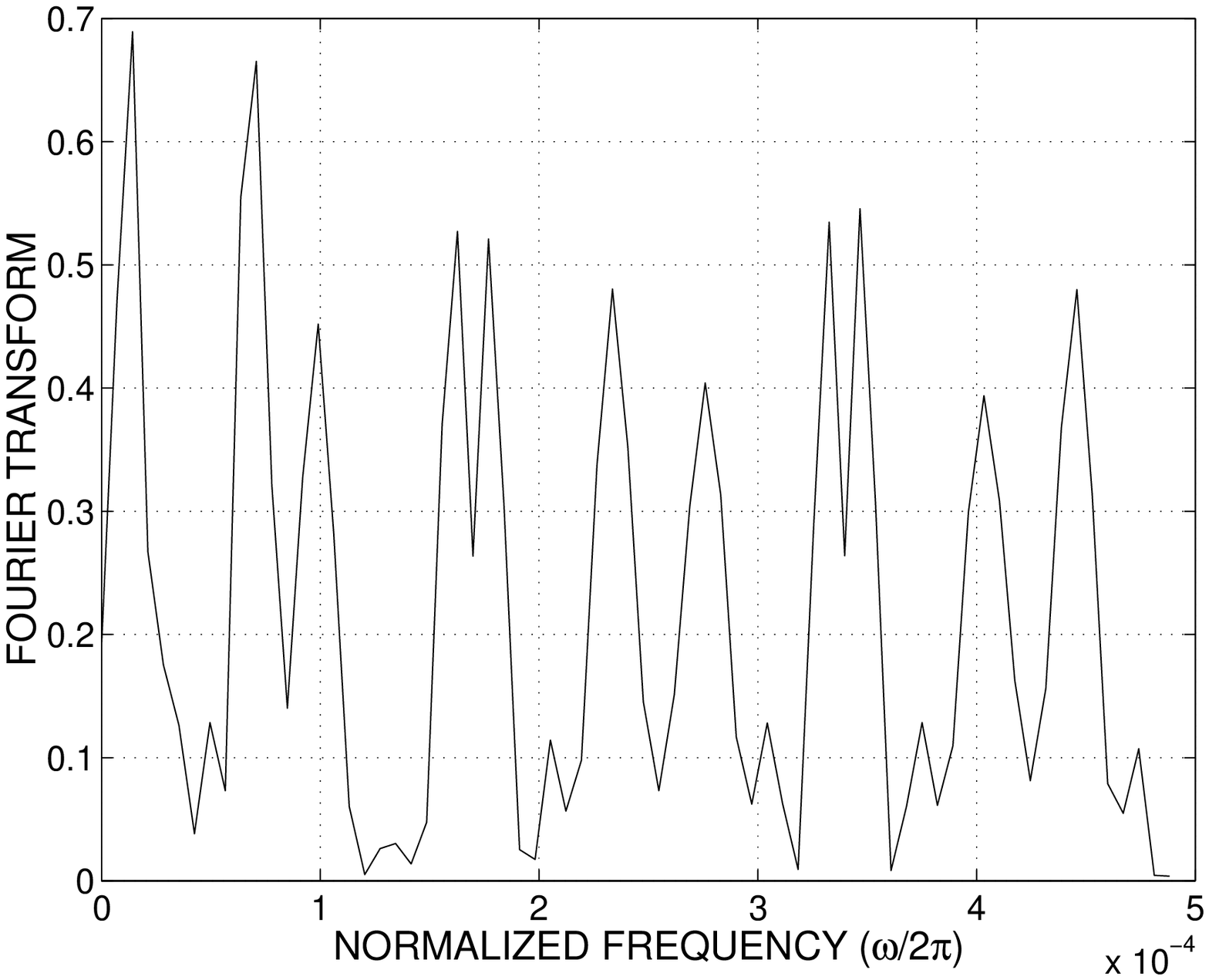,height=3cm}
}
\caption{Left : $\dtft{q}(\dfv)$.
Right : $\dtft{q}(\dfv)$ of the corrupted FIF.
}
\label{corruptfif}
\end{figure}

The above algorithm was tested in FIF signals contaminated with
white noise and the parameters extracted were very close to the
real, fig.~\ref{noisy_fif}. The main advantage of the method is
the use of FFT. It is well known that FFT is quite simple and
easily implemented. The first disadvantage of the method is that
it cannot find the FIF order. In the above experiments we tried
for $N=2,3,\dots$ until the estimated ${\dtft q}$ function
satisfied the periodicity and symmetry conditions mentioned
earlier. The second and most important problem is that it is very
sensitive in the window effect of the FFT. It is known that
although a part of FIF signal is self affine it is not an FIF.
Trying to model it as an FIF results in wrong estimations. In the
example of fig.~\ref{corruptfif}, it  is evident that the period
of the ${\dtft q}$ function has been expanded. Having chosen the
order $N$, the period of ${\dtft q}$ is known $2\pi
\frac{N}{M-1}$. In the right plot of fig.~\ref{corruptfif} it is
evident that the period of ${\dtft q}$ is much higher than the
expected. But although the spectrum is not the best method for
solving the inverse problem it can be used as powerful analysis
tool. As shown above it can reveal the FIF nature of a signal and
it can also help in the prediction of a missing part of a time
series, given that it belongs to class of FIF.

\begin{thebibliography}{9}


\bibitem{mandelbrot} B. B. Mandelbrot, "The Fractal Geometry of Nature"
W.H Freeman and Company, New York, 1983.

\bibitem{barnsley} M. F. Barnsley,
"Fractals Everywhere",
Academic Press, 1993.

\bibitem{mazel} D. S. Mazel, M. H. Hayes,
"Using Iterated Function Systems to Model Discrete Sequences",
{\em IEEE Transactions on Signal Processing}, Vol. 40 (7), pp.~1724-1734, July 1992.

\bibitem{abarbanel} H. D.I. Abarbanel,
"Analysis of Observed Chaotic Data",
Springer Verlag, New York 1996.

\bibitem{oppenheim} A. V.Oppenheim, R. W. Schafer,
"Deiscrete-Time Signal Processing",
Prentice Hall, 1999.
\end{thebibliography}
\end{document}